

\documentstyle{amsppt}
\TagsOnRight


\nopagenumbers

\input amstex
\magnification=\magstep1

\def\ad{\operatorname{ad}}
\def\+{\oplus }
\def\x{\times }

\def\dm{-\cdots -}

\def\h{\calH}

\def\g{\calG}
\def\dgae{differential graded Lie algebra}
\def\dga{\dgae\ }
\def\k{\calK}
\def\f{\Phi}
\def\p{\Psi}

\def\e{\eps}
\def\r{\rho}

\def\mod{\operatorname{mod\ }}
\def\hl{\hatL}
\def\ha{\hatA}
\def\hb{\hatB}
\def\d{\del}
\def\slim{\operatornamewithlimits{s-lim\ }}

\def\nra{NR1}
\def\nrb{NR2}
\def\kura{Ku1}
\def\kurb{Ku2}
\def\ksa{KS}
\def\mk{MK}
\def\fn{FN}
\def\kns{KNS}
\def\gm{GM1}
\def\gmb{GM2}
\def\ar{A}
\def\kato{Ka}

\document

\topmatter

\title
Deformation of Infinite Dimensional
Differential Graded Lie Algebras
\endtitle
\author
Maxim Braverman
\endauthor
\address
School of Mathematical Sciences,
Tel-Aviv University,
Ramat-Aviv 69978, Israel
\endaddress

\abstract
{}For a wide class of topological infinite dimensional \dgae s the
complete set of deformations is constructed. It is shown that
for several deformation
problems the existence of a formal power series solution guarantees
the existence
of an analytic solution.
\endabstract

\thanks
  The research was supported by grant No. 449/94-1 from the
  Israel Academy of Sciences and Humanities.
\endthanks

\email
maxim\@math.tau.ac.il
\endemail

 \endtopmatter

\def\today{\ifcase\month\or January\or February\or March\or
April\or May\or June\or July\or August\or September\or
October\or November\or December\fi \space\number\day,
\number\year}

\def\alp{\alpha}		
\def\bet{\beta}
		
\def\del{\delta}		
\def\eps{\varepsilon}


\def\calG{{\Cal G}}
\def\calH{{\Cal H}}

\def\calK{{\Cal K}}




\font\tenboldgreek=cmmib10  \font\sevenboldgreek=cmmib10 at
7pt
\font\fiveboldgreek=cmmib10 at 7pt
\newfam\bgfam
\textfont\bgfam=\tenboldgreek \scriptfont\bgfam=\sevenboldgreek
\scriptscriptfont\bgfam=\fiveboldgreek


\font\tengerman=eufm10 \font\sevengerman=eufm7 \font\fivegerman=eufm5
\font\tendouble=msym10 \font\sevendouble=msym7 \font\fivedouble=msym5

\textfont4=\tengerman \scriptfont4=\sevengerman
\scriptscriptfont4=\fivegerman
\newfam\dbfam
\textfont\dbfam=\tendouble \scriptfont\dbfam=\sevendouble
\scriptscriptfont\dbfam=\fivedouble

\mathchardef\ng="702D
\mathchardef\dbA="7041
\mathchardef\sm="7072
\mathchardef\nvdash="7030
\mathchardef\nldash="7031
\mathchardef\lne="7008
\mathchardef\sneq="7024
\mathchardef\spneq="7025
\mathchardef\sne="7028
\mathchardef\spne="7029
\mathchardef\ltms="706E
\mathchardef\tmsl="706F

\mathchardef\dbA="7041



\def\RR{\Bbb R}


\def\nek{,\ldots,}
\def\sdp{\times \hskip -0.3em {\raise 0.3ex
\hbox{$\scriptscriptstyle |$}}} 


\def\End{\operatorname{End\,}}

\def\IM{\operatorname{Im}}








\def\hatA{{\widehat A}}

\def\hatB{{\widehat B}}

\def\hatL{{\widehat L}}



\def\tilf{{\widetilde f}}

\def\tilm{{\widetilde m}}


\def\sqr#1#2{{\vcenter{\hrule height.#2pt\hbox{\vrule
width.#2pt height#1pt \kern#1pt \vrule width.#2pt}\hrule
height.#2pt}}}

\def\buildrul#1\under#2{\mathrel{\mathop{\null#2}\limits_{#1}}}

\def\boxit#1{\vbox{\hrule\hbox{\vrule\kern3pt\vbox{\kern3pt#1
\kern3pt}\kern3pt\vrule}\hrule}}

\def\subheading#1{\medskip\goodbreak\noindent{\bf
#1.}\quad}

\def\longmapright #1 #2 {\smash{\mathop{\hbox to
#1pt {\rightarrowfill}}\limits^{#2}}}
\def\longmapleft #1 #2 {\smash{\mathop{\hbox to
#1pt {\leftarrowfill}}\limits^{#2}}}

\def\back{{\raise 2.5pt\hbox{$\,\scriptscriptstyle\backslash\,$}}}

\def\part{\partial}
\def\lwr #1{\lower 5pt\hbox{$#1$}\hskip -3pt}
\def\rse #1{\hskip -3pt\raise 5pt\hbox{$#1$}}
\def\lwrs #1{\lower 4pt\hbox{$\scriptstyle #1$}\hskip -2pt}
\def\rses #1{\hskip -2pt\raise 3pt\hbox{$\scriptstyle #1$}}

\def\<#1{\left\langle{#1}\right\rangle}

\def\subinbn{{\subset\hskip-8pt\raise
0.95pt\hbox{$\scriptscriptstyle\subset$}}}

\def\llvdash{\mathop{\|\hskip-2pt \raise 3pt\hbox{\vrule
height 0.25pt width 1.5cm}}}

\def\lvdash{\mathop{|\hskip-2pt \raise 3pt\hbox{\vrule
height 0.25pt width 1.5cm}}}

\def\fakebold#1{\leavevmode\setbox0=\hbox{#1}%
  \kern-.025em\copy0 \kern-\wd0
  \kern .025em\copy0 \kern-\wd0
  \kern-.025em\raise.0333em\box0 }

\font\msxmten=msxm10
\font\msxmseven=msxm7
\font\msxmfive=msxm5
\newfam\myfam
\textfont\myfam=\msxmten
\scriptfont\myfam=\msxmseven
\scriptscriptfont\myfam=\msxmfive
\mathchardef\rhookupone="7016
\mathchardef\ldh="700D
\mathchardef\leg="7053
\mathchardef\ANG="705E
\mathchardef\lcu="7070
\mathchardef\rcu="7071
\mathchardef\leseq="7035
\mathchardef\qeeg="703D
\mathchardef\qeel="7036
\mathchardef\blackbox="7004
\mathchardef\bbx="7003
\mathchardef\simsucc="7025

\def\+{\oplus }
\def\x{\times }

\def\dm{-\cdots -}

\heading 0. Introduction \endheading

\subheading{0.1}In \cite\nrb\ Nijenhuis and Richardson
developed deformation
theory of finite dimensional \dgae s.
In particular, for a \dga $(L,d)$ they
constructed a versal deformation, which is a precise analogue
of the versal
deformation of a compact complex manifold constructed by Kuranishi
(\cite\kura, \cite\kurb). The parameter space of this deformation is a
germ of a finite dimensional analytic variety. This germ is
called {\it the Kuranishi space} of $L$.

\subheading{0.2}In \cite\gmb\ Goldman and Millson
defined a notion of   {\it
analytic \dgae}. Namely, a  \dga $(L=\bigoplus L^i,d)$ is analytic if
\roster
\item it has finite cohomology in degrees 0 and 1,
\item each $L^i$ is a normed vector space,
\item the maps $d:L^i\to L^{i+1}$, and \,
	$\ad a:L^i\to L^{i+1}\ (a\in L^1)$ are continuous,
\item  the completion $\hl^i$ of $L^i$ may be
	decomposed into a direct sum of
closed subspaces
$$\hl^i=\hb^i\+\h^i\+\ha^i,$$
where  \,$\hb^i$ is the image of $d$ in $\hl^i$,
\,$\h^i+\hb^i$ is the kernel of $d$ in $\hl^i$ \,and \,$\h^i\i L^i$.
\endroster
For an analytic \dga Goldman and Millson constructed the
Kuranishi space, which
is a germ of a finite dimensional analytic variety,
as in the case of finite dimensional \dgae s.
It is shown in \cite\gmb\ that the Kuranishi space is
invariant under quasi-isomorphisms of \dgae s.

\subheading{0.3}For a general analytic \dga
it is impossible to define the
equivalence of deformations and the
Kuranishi space may be considered as a
complete set of
deformations only formally.

In this paper we define an {\it elliptic \dgae} to be an
analytic \dga such that for any element \,$a\in \hl^0$ \,the operator \
$\ad a:\hl^1\to\hl^1$ generates a one parameter semigroup of bounded
operators \ $\exp(t\ad a):\hl^1\to\hl^1$. The class of elliptic \dgae s
contains  such  examples as the finite dimensional \dgae,
the  twisted de Rham algebra of differential forms with
values in endomorphisms of a flat
vector bundle over a compact manifold, etc.

\subheading{0.4}By an ($n$ - parameter)
{\it analytic deformation\/} of a \dga
$(L,d)$ we shall mean (cf. \cite\nrb) an analytic map
$$m:U\to L^1, \qquad u\mapsto \ m(u)$$
defined on a neighborhood\  $U$ of $0$ in
$\RR^n$, such that $m(0)=0$ and
$$dm(u)+\frac12[m(u),m(u)]=0.$$
For each $a\in \hl^0$ we define the  operator \
$\r(a):\hl^1\to\hl^1$ by
the formula
$$
\r(a):x\mapsto \exp(\ad a)\,x-\left[\int_0^1\exp((1-t)\ad a)
\,dt\right]\,da.
$$

\subheading{0.5}We shall say that two analytic deformations
$m(u)\ (u\in U\i\RR^n)$ and $m'(u)\ (u\in U'\i\RR^n)$
are  {\it equivalent} if there exists a neighborhood
of zero $V\i U\cap U'$
such that for any $v\in V$ there exists $f\in \hl^0$ such that
$\r(f)\,m(v)=m'(v)$.
If the element $f$ may be chosen so that $f\in\ha^0$
we shall say that the deformations $m(u)$ and $m'(u)$ are
{\it similar}.


\subheading{0.6}An analytic deformation \
$m:U\to L^1\ (0\in U\i \RR^n)$
will be called a
{\it Kuranishi deformation} if \ $m(u)\in \h^1\+\ha^1$ provided $u$ is
small enough.

The following proposition  generalizes a  result of
Nijenhuis and Richardson (\cite\nrb) :

\proclaim{Proposition 1}
Any analytic deformation of an elliptic \dga is similar
to a unique Kuranishi deformation.
\endproclaim

\subheading{0.7}We use the above result to show that
for some deformation
problems the existence of a formal power series solution guarantees the
existence of an analytic solution. More precise, we define  a
{\it formal deformation} of \dga $(L,d)$ to be a formal power series
$$m(t)=tm_1+t^2m_2+\cdots$$
with coefficients $m_n\ (n=1,2,\dots)$ in $L^1$ satisfying
$$
dm(t)+\frac12[m(t),m(t)]=0.
$$
(For simplicity we consider only 1-parameter formal deformations.
Generalizations to the $n$-parameter case are immediate). Two formal
deformations $m(t)$ and $m'(t)$ will be called
{\it formally similar} if there exists a formal power series
$$f(t)=tf_1+t^2f_2+\cdots$$
with coefficients $f_n\ (n=1,2,\dots)$ in $\ha^0\i \hl^0$ such that
$$
m'(t)=\r(f(t))\,m(t)\overset\text{def}\to
=\exp(\ad f(t))\,m(t)+\frac{I-\exp(\ad f(t))}{\ad f(t)}(df(t)). $$
Our main results are the following two theorems:
\proclaim{Theorem 2}
If two analytic deformations of an elliptic \dga are
formally similar, then they are similar.
\endproclaim
\proclaim{Theorem 3} For any formal deformation
$m(t)$ of an elliptic \dga
$(L,d)$ and any integer $n$ there exists an analytic
deformation $\tilm(t)$
such that

$$\tilm(t)\equiv m(t)\qquad \mod t^n.$$
\endproclaim
Note that in finite dimensional case the above
theorems follow directly from
Artin's result (\cite\ar).

\heading Acknowledgments \endheading

This work was started together with  M. Farber and S. Shnider.
Therefore I would
like first to thank them for all they taught me during this work,
and, second,
to chide them for steadfastly refusing to share the blame
with me as joint
authors.

I am very thankful to V. Matsaev for valuable discussions and to
M. Kontsevich for pointing out a gap in the preliminary version
of the paper.

\heading  1. Elliptic Differential Graded Lie Algebras \endheading

In this section we define the notion of elliptic \dgae,
which will be the main object of study in the present paper.

\subheading{1.1}A {\it graded Lie algebra}
(\cite\nrb, \cite\gm) is  a complex vector
space
$L=\bigoplus_{0\le i\le N}L^i$
graded by a nonnegative integers, and a family of bilinear maps
$$[\cdot,\cdot]:L^i\x L^j\to L^{i+j}$$
satisfying (graded) skew-commutativity:
$$[a,b]+(-1)^{ij}[b,a]=0$$
and  the (graded) Jacobi identity:
$$[a,[b,c]]=[[a,b],c]+(-1)^{ij}[b,[a,c]]$$
where $a\in L^i,\ b\in L^j,\ c\in L^k$.

\subheading{1.2}A {\it differential graded Lie algebra}
is a pair $(L,d)$ where $L$ is a graded
Lie algebra and $d$ is a family of linear maps
\,$d:L^i\to L^{i+1}\ (0\le i\le N)$ such that \,$d\circ d=0$ and
$$d[a,b]=[da,b]+(-1)^i[a,db]$$
where \,$a\in L^i,\ b\in L$.


\subheading{1.3}A {\it normed \dga} (\cite\gmb) is a \dga
$(L,d)$ equipped with a norm $|\cdot|_i$ on each $L^i$ making
$L^i$ into a normed vector space such that the maps
$$d:L^i\to L^{i+1}$$
and
$$\ad a:L^i\to L^{i+1},\ \qquad a\in L^1$$
are continuous.

Note that in \cite\gmb\ the map $\ad a:L^i\to L^{i+1}\ \
a\in L^0$ is demanded
to be continuous only for $i=1$.

Let $\hl^i$ be the completion of $L^i$ for $i=0,1,2,\dots$.
The vector space $L^i$ is the analogue of the smooth $i$-forms
and $\hl^i$ of the Sobolev $i$-forms.

\subheading{1.4}An {\it analytic \dgae} (\cite\gmb) is a normed
\dga with finite dimensional cohomology in degree 0 and 1,
equipped with  decompositions of each $\hl^i$ into
direct sums of closed subspaces:
$$\hl^i=\hb^i\+\h^i\+\ha^i,
\tag1.1$$
where \,$\hb^i$ is the image of $d$ in $\hl^i$,
\,$\h^i+\hb^i$ is the kernel of $d$ in $\hl^i$ \,and \,$\h^i\i L^i$.
Let \,$\bet:\hl^i\to \hb^i,\ \alp:\hl^i\to \ha^i$
and \,$H:\hl^i\to \h^i$ \,be the corresponding projections.

We assume that all three projections carry $L^i$ into itself and that
$\bet(L^i)=dL^{i-1}$.
Set  \,$A^i=\ha^i\cap L^i$ \,and \,$B^i=\hb^i\cap L^i$.
Obviously, $A^i=\alp(L^i), \ B^i=\bet(L^i)$
\,and there is an algebraic direct sum decomposition
$$L^i=B^i+\h^i+A^i. \tag1.2$$
There exists (\cite\gmb) a continuous operator
\,$\d:\hl^{i+1}\to \hl^i$
\,such that \,$\d(L^{i+1})\i L^i$ \,$\ha^i=\IM\d$ \,and
$$id =H+d\circ\d+\d\circ d. \tag1.3$$


\subheading{1.5}We remind (\cite\kato\ Ch. IX)
that  a family of bounded linear operators \,$U(t):\hl^1\to\hl^1$
\,depending on a parameter $t\ (0\le t<\infty)$
is called a quasi-bounded one-parameter semigroup if  $U(0)=id$,
$$U(t_1+t_2)=U(t_1)U(t_2) \qquad (0\le t_1,t_2 <\infty),$$
and there exist numbers $C>0$ and $\bet$ such that
$$\|U(t)\|\le Ce^{\bet t}\qquad  ( 0\le t<\infty).
\tag1.4$$
In this case the strong limit
$$T=\slim_{t\to+0}\frac{U(t)-id}{t}$$
exists and is a densely defined operator $\hl^1\to\hl^1$. We
shall say that $T$ generates the semigroup
$U(t)$ and write $U(t)=\exp(tT)$.

The description of the class of operators which generate a
quasi-bounded one-parameter semigroup may be found in
\cite\kato\ Ch. IX \S 1.4.
Note that any bounded operator belongs to this class.

\subheading{1.6. Definition}By an
{\it elliptic \dga} we shall understand an analytic
\dga $(L,d)$ such that
for each $a\in \hl^0$ the operator $\ad a$ generates
a quasi-bounded one-parameter semigroup of operators
$\exp(t\ad a):\hl^1\to\hl^1$ depending continuously on
$a\in \hl^0$, i.e. for any $\eps>0$ there exists a
neighborhood $U$ of 0 in $\hl^0$ such that for any
$a\in U$ and for any $0\le t\le 1$
$$\|\exp(t\ad a)-id\|<\eps.  \tag1.5$$
Here $\|\cdot\|$ denotes the standard norm on
the space of operators acting on
the Banach space $\hl^1$.

\subheading{1.7. Remark} In many examples (cf. section 1.8) the
operators $\ad a:\hl^1\to \hl^1\ (a\in \hl^0)$
are bounded. In this case
$\exp(t\ad a)$ is given by a convergent power series
$$
   \exp(t\ad a)=\sum_{i=0}^\infty \frac{t^i}{i!}(\ad a)^i.
$$
Hence any analytic \dga of this type is elliptic. In particular,
any finite dimensional \dga is elliptic.

We finish this section with an  example of an infinite dimensional
\dgae.
\subheading{1.8. The twisted de Rham algebra}Let $M$ be a compact
$n$-dimensional manifold
and $\Cal E$ a flat vector bundle over $M$.
The vector bundle $End(\Cal E)$ inherits the flat structure from
$\Cal E$.
The {\it twisted de Rham algebra} is the \dga
$$(L,d)=\Big(\bigoplus_{i=0}^n A^i(M,End(\Cal E)), d\Big)$$
of all forms on $M$ with
values in the bundle $End(\Cal E)$.

Fix $s\in\RR$ and equip the space
$A^i(M,\End(\Cal E))\ (0\le q\le n)$ with
$(s-q)$-th Sobolev's norm. Then,
by the classical Hodge theory,  the algebra $(L,d)$ is
analytic. Since the map
$$\gather
A^0(M,End(\Cal E))\x A^1(M,End(\Cal E))\to A^1(M,End(\Cal E)),\\
(a,x)\mapsto [a,x]
\endgather$$
is continuous this algebra is elliptic (cf. Remark 1.7).


\heading  2. Deformations of Elliptic Differential Graded Lie Algebras
\endheading

In this section we define the notions of analytic deformations of
elliptic
\dgae s and of analytic equivalence of two analytic deformations.
\subheading{2.1}Let $(L,d)$ be an elliptic \dgae. The equation
$$da+\frac12[a,a]=0, \qquad a\in L^1\tag2.1$$
is called a {\it deformation equation\/} (\cite\nrb). We denote by
$M\i L^1$ the set of solutions of (2.1).
For any \,$m\in M$ \,the derivation
\,$d_m=d+\ad m:L\to L$ \,satisfies the equation
$d_m\circ d_m=0$.  Hence
the \dga $(L,d_m)$ is defined.

By an ($n$ - parameter) {\it analytic deformation\/} of a
\dga $(L,d)$ we shall mean an analytic map
$$
m:U\to M, \qquad        u\mapsto m(u)
$$
defined on a neighborhood $U$ of $0$ in $\RR^n$ such that $m(0)=0$.
The trivial map $m(u)\equiv 0$ defines the {\it trivial deformation\/}.

\subheading{2.2}Fix $a\in\hl^0, \ x_0\in\hl^1$ and let $x(t)$
denote the solution of the inhomogeneous differential equation
$$
\frac{dx}{dt}=[a,x]-da, \qquad x(0)=x_0 .
\tag2.2$$
Obviously,
$$
d+\ad x(t)=\exp(t\ad a)\circ \big(d+\ad x(0)\big)\circ\exp(-t\ad a).
$$
We shall denote by $\r(a)$ the map $\r(a):x_0\mapsto x(1)$.
By \cite\kato\ Ch. IX \S1.5 \ $\r(a)$ is an affine transformation
of $\hl^1$
given by the formula
$$
\r(a):x_0\mapsto
\exp(\ad a)\,x_0-\left[\int_0^1\exp((1-t)\ad a)\,dt\right]\,da.
\tag2.3$$
Formally one can rewrite (2.3) as (cf. \cite\gm\ \S 1)
$$\r(ta)\,x=\exp(t\ad a)\,x+\frac{I-\exp(t\ad a)}{\ad a}(da). \tag2.4$$


\subheading{2.3}Let $\g$ denote the group of affine transformations
of $\hl^1$ generated by the operators $\r(a)$ with $(a\in \hl^0)$.

Two elements $m,m'\in M$ are called {\it equivalent} if there
exists $g\in \g$ such that $g(m)=m'$.
If the element $g$ may be chosen so that  $g=\r(f),\
f\in\ha^0$ then the elements $m$ and $m'$ are called {\it similar}.
Note that the similarity is not an equivalence relation.

Two analytic deformations $m(u)\ (u\in U\i\RR^n)$ and
$m'(u)\ (u\in U'\i\RR^n)$
are called {\it equivalent\/} (respectively, {\it similar})
if there exists a neighborhood of zero $V\i U\cap U'$
such that for any $v\in V$ the elements $m(v)$ and $m'(v)$
are equivalent (respectively, similar).

\heading 3. The Kuranishi Space \endheading

In this section we shall extend the Kuranishi's theory of locally
complete families of complex structures (\cite\kura, \cite\kurb)
to  arbitrary elliptic differential graded Lie algebras.
For finite dimensional \dga this theory was developed by
Nijenhouis and Richardson (\cite\nra, \cite\nrb).
In \cite\gmb\ the formal analog of this theory was developed
for an analytic \dgae s.

\subheading{3.1}The {\it Kuranishi} map
$F:\hl^1\to \hl^1$ is a quadratic map defined by
$$F(a)=a+\frac12\d[a,a].$$
We observe that $F(L^1)\i L^1$. It is shown in \cite\gmb\
(Lemma 2.2) that there exist balls $B$ and $B'$ around $0$
in $\hl^1$ such that $F$ is an analytic diffeomorphism  $B\to B'$.
The {\it Kuranishi space} $\k\i\h^1$ is defined by
$$\k=\Big\{a\in B'\cap\h^1:\ H\big[F^{-1}(a),F^{-1}(a)\big]=0\Big\}.$$
Let
$$Y=\big\{a\in \h^1\+\ha^1: da+\frac12[a,a]=0\big\}.$$
The following theorem is proven in \cite\gmb\ (Theorem 2.3)


\proclaim{3.2. Theorem} $F$ induces a homeomorphism from a
neighborhood of $0$ in $Y$ to a neighborhood of $0$ in $\k$.
\endproclaim
Since $\k\i\h^1\i L^1$ and $F(L^1)\i L^1$ it follows from this
theorem that $Y\i L^1$.

\subheading{3.3}We shall say that the set $X\i M$ is a
{\it locally complete set of deformations} if there exists a
neighborhood $U\i M$ of the origin such that any $m\in U$
is equivalent to some element of $X$. Kuranishi
(\cite\kura,\cite\kurb) showed that in the case of the
Kodaira-Spencer algebra the set $Y$ is locally complete.
We shall show that it is  also true for an elliptic \dgae.
In fact, we need a bit stronger result.


\proclaim{3.4. Lemma}Let $(L,d)$ be an elliptic \dgae.
Then there exist a neighborhood $U$ of $0$ in $M$, a neighborhood
$V$ of $0$ in $\hl^0$ and a neighborhood $W$ of 0 in $\ha^0$
such that for any  $m\in U$ and any $s\in V$ there exists a
unique   element $f\in W$ such that
$$\r(s)\circ\r(f)\,(m)\in Y.$$
\endproclaim

\demo{Proof}Let $\f:\hl^1\x\hl^0\x\hl^0\to\hl^1$ be the map
defined by the formula:
$$
\f(x,a,s)= \r(s)\circ\r(a)\,(x)-x+da+
\left[\int_0^1\exp((1-t)\ad s)\,dt\right]\,ds,
\tag3.1$$
where $x\in L^1$ and $a\in L^0$.
We need to find  $f$ and $m'$\  such that
$$
\gathered
m'=\r(s)\circ\r(f)\,(m),\\
f\in \ha^0,\qquad
m'\in\h^1\+\ha^1.
\endgathered
\tag3.2$$
Denote $l=df+m'$. Then (3.2) is equivalent  to
$$\gathered
l=m-\left[\int_0^1\exp((1-t)\ad s)\,dt\right]\,ds+\f(m, \d l,s),\\
m'=(H+\d d)l,\qquad
 f=\d l .
\endgathered \tag3.3$$
It follows from (2.3) and (1.5) that there exist a neighborhood
$U$ of $0$ in $M$, a neighborhood $V$ of $0$ in $\hl^0$ and a
neighborhood $W$ of 0 in $\ha^0$ such that for any  $m\in U$,
any $s\in V$ and any $l,l'\in W$
$$m-\left[\int_0^1\exp((1-t)\ad s)\,dt\right]\,ds+\f(m, \d l,s)\in W$$
and
$$|\f(m,\d l, s)-\f(m,\d l', s)|_1< \frac12|l-l'|_1.$$
(Remind that by $|\cdot|_1$ we denote the norm on $L^1$).
Then for any $m\in U, s\in V$ the equation (3.3) has a unique
solution $l\in W$ given, say, by iterations
$$\gather
l^{(r+1)}=
m-\left[\int_0^1\exp((1-t)\ad s)\,dt\right]\,ds+\f(m,\d l^{(r)},s),\\
l^{(1)}=m-\left[\int_0^1\exp((1-t)\ad s)\,dt\right]\,ds.
\endgather
$$
The lemma is proved.
$\square$

\enddemo

\subheading{3.5}An analytic deformation $m:U\to M$\
$(0\in U\i\RR^n)$ of an elliptic \dga is called a
{\it Kuranishi deformation} if $m(u)\in Y$ for any $u\in U$.

Note that the element $f\in \hl^0$ defined by (3.3) depends
analytically on $m$.
Hence, we obtain the following


\proclaim{3.6. Corollary} Let \,$m:U\to M$ \,$(0\in U\i\RR^n)$
\,be an analytic deformation of an elliptic \dga $(L,d)$ and let
\,$s:U\to L^0$  \,be an analytic map. Then there exist a neighborhood
\,$V\i U$ of \,$0$ in $\RR^n$ and a unique analytic function
\,$f:V\to \ha^0$ \,such that $f(0)=0$ and
$$\r(s(u))\circ\r(f(u))\,(m(u))\in Y,\qquad u\in V.$$

\endproclaim

Setting in the above lemmas $s\equiv 0$ we obtain the following:

\proclaim{3.7. Proposition}  1. Let $(L,d)$ be an elliptic \dgae.
Then there exists a neighborhood $U$ of $0$ in $M$ such
that any element $m\in M$ is similar to a unique $m'\in Y$.
In particular,
$Y=\big\{a\in \h^1\+\ha^1: da+\frac12[a,a]=0\big\}$
is a locally complete set of deformations.

2.  Any analytic deformation of an elliptic \dga is similar
to a unique Kuranishi deformation.

\endproclaim

\heading  4. Formal Theory \endheading

In this section we shall show that for some
deformation problems the existence
of a formal power series solution guarantees the existence of
an analytic solution.

\subheading{4.1}By a (1-parameter) {\it formal deformation} of a
\dga $(L,d)$ we shall understand a formal power series
$$
m(t)=tm_1+t^2m_2+\cdots$$
with coefficients $m_n\ (n=1,2,\dots)$ in $L^1$ satisfying
$$
dm(t)+\frac12[m(t), m(t)]=0.
\tag 4.1$$
Note that any analytic deformation may be considered as a formal one.

Two formal deformations $m(t)$ and $m'(t)$ are called
{\it formally equivalent} if there exists a formal power series
$$f(t)=tf_1+t^2f_2+\cdots$$
with coefficients $f_n\ (n=1,2,\dots)$ in $\hl^0$ such that (cf. (2.4))
$$
m'(t)= \r(f(t))\,m(t)\overset\text{def}\to=
\exp(\ad f(t))\,m(t)+\frac{I- \exp(\ad f(t))}{\ad f(t)}(df(t)).
\tag 4.2$$
If $f(t)$ in (4.2) may be chosen so that
$$f_n\in\ha^0
\tag 4.3$$
for any $n=1,2,\dots$, we shall say that the deformations
$m(t)$ and $m'(t)$ are {\it  formally similar}.

A formal deformation $m(t)=tm_1+t^2m_2+\cdots$ is called a
{\it formal Kuranishi deformation} if
$$m'_n\in\h^1\+\ha^1\tag 4.4$$
for any $n=1,2,\dots$.

\proclaim{4.2. Lemma} Let $m(t)=tm_1+t^2m_2+\cdots$ be a formal
deformation of an elliptic \dga $(L,d)$ and let
$s(t)=ts_1+t^2s_2+\cdots$ be a power series with coefficients in
$\hl^0$. Then there exist a unique formal Kuranishi deformation
$m'(t)=tm_1'+t^2m_2'+\cdots$ and a unique power series
$f(t)=tf_1+t^2f_2+\cdots$ with coefficients in $\ha^0$ such that
$$m'(t)= \r(s(t))\circ \r(f(t))\,(m(t)). \tag4.5$$
\endproclaim
\demo{Proof}Let \,$\p:\hl^1\x \hl^0\x \hl^0\to \hl^1$
\,be the  map defined by
the formula
$$
\p(x,a,s)=\r(s)\circ\r(a)(x)-x+da+ds.
$$
The formal power series $f(t)$ and $m'(t)$ satisfy (4.3),(4.4) and
(4.5) if and only if the formal power series
$l(t)=tl_1+t^2l_2+\cdots=df(t)+m'(t)$ satisfies the equation:
$$
l(t)=m(t)-ds(t)+\p(m(t),\d l(t),s(t)).
\tag 4.6$$
As the formal power series \,$\p(m(t),\d l(t),s(t))$ \,does
not contain summands of degree $<2$ in $t$, the equation (4.6)
defines the coefficients \,$l_n$ of \,$l(t)$ recursively
$$
l_n=l_n(m_1\nek m_n;l_1\nek l_{n-1};s_1\nek s_n),\qquad n=1,2,\dots
$$
Hence, there exists a unique formal solution $l(t)$ of (4.5). $\square$
\enddemo

The following corollary  is the formal analogue of proposition 3.7.
\proclaim{4.3. Corollary} Let $m(t)=m_1+t^2m_2+\cdots$ be a formal
deformation of an elliptic \dga $(L,d)$. Then there exists a
unique formal Kuranishi deformation $m'(t)=m_1'+t^2m_2'+\cdots$
which is formally similar to $m(t)$.
\endproclaim

The above lemma together with corollary 3.6 implies the following
\proclaim{4.4. Theorem} If two analytic deformations of an
elliptic \dga are  formally similar, then they are similar.
\endproclaim
\demo{Proof}Let analytic deformations $m'(t)$ and $m''(t)$ be
formally similar and let $f(t)=tf_1+t^2f_2+\cdots$ be the formal
power series with coefficients in $\ha^0$ such that
$m'(t)= \r(f(t))\,(m''(t))$. We shall show that the power
series $f(t)$ converges in a neighborhood of $0$.

By proposition 3.7 there exists $\e>0$ and an analytic function
$s(t)\in \hl^0$ defined for $t\in(-\e,\e)$ such that  $s(0)=0$ and
$$
 \r(s(t))\,(m'(t))\in Y,\qquad t\in(-\e,\e).
$$
Then
$$ \r(s(t))\circ \r(f(t))\,\big( m''(t)\big)$$
is a formal Kuranishi deformation.

By corollary 3.6 there exists \,$0<\del<\e$ \,and an analytic
function \,$\tilf(t)\in\ha^0$ defined for
$t\in(-\del,\del)$ such that
$$
 \r(s(t))\circ \r(\tilf(t))\,\big( m''(t)\big)\in Y,
\qquad t\in(-\del,\del).
$$

The uniqueness statement in lemma 4.2  implies $f(t)\equiv\tilf(t)$.
$\square$
\enddemo


\proclaim{4.5. Theorem}For any formal deformation $m(t)$ of
an elliptic \dga $(L,d)$ and any integer $n$ there exists an
analytic deformation $\tilm(t)$ such that
$$\tilm(t)\equiv m(t)\qquad \mod t^n.
\tag 4.7$$
\endproclaim
\demo{Proof}Let $m'(t)$ be the unique formal Kuranishi
deformation equivalent to
 $m(t)$ and let $f(t)=tf_1+t^2f_2+\cdots$ be a formal
power series with coefficients
$f_n\in \hl^0$ such that
$$ \r(f(t))\,m(t)=m'(t).$$
Denote
$$m'_*(t)=tHm'_1+t^2Hm'_2+\cdots.$$
Then $m'_*(t)$ is a formal power series with coefficients in the
finite dimensional space $\h^1$.  By theorem 3.2 the series
$m'(t)$ is a formal Kuranishi deformation if and only if
$m'_*$ satisfies
the system of analytic equations
$$H[F^{-1}(m'_*),F^{-1}(m'_*)]=0.\tag 4.8$$
By Artin's theorem \cite\ar\ there exists an analytic
solution $m_*(t)\in \h^1$ of (4.8) such that
$$m_*(t)\equiv m_*'(t)\qquad \mod t^n.$$
Denote \,$m''(t)=F^{-1}(m_*(t))$.
Then $m''(t)$ is an analytic Kuranishi deformation. The deformation
$$
\tilm(t)= \r(-tf_1-t^2f_2\dm t^{n-1}f_{n-1})\,(m''(t))
$$
is an analytic deformation and satisfies (4.7). $\square$
\enddemo

\enddocument